\documentclass{aa}
\usepackage{psfig,times}
\usepackage{amssymb}
\usepackage{graphicx}

\begin{document}
\title{Protoneutron star dynamos and pulsar magnetism}

\author{A.~Bonanno$^{1,2}$, V.~Urpin$^{3,4}$, and G.~Belvedere$^{1}$}
\offprints{A.Bonanno}

\institute{$^{1}$ INAF, Osservatorio Astrofisico di Catania,
        Via S.Sofia 78, 95123, Catania, Italy \\
        $^{2}$ INFN, Sezione di Catania, 
	Via S.Sofia 72, 95123, Catania, Italy \\
	$^{3}$ A.F. Ioffe Institute of Physics and Technology, 
	St. Petersburg, Russia \\
	$^{4}$ Isaac Newton Institute of Chile, Branch in St. Petersburg, 
        194021 St. Petersburg, Russia\\
}

\titlerunning{Protoneutron star dynamos}
\authorrunning{Bonanno, Urpin, and Belvedere}
\date{\today}

\abstract{We have investigated the turbulent mean-field dynamo action in
          protoneutron stars that are subject to convective and neutron
	  finger instabilities during the early evolutionary phase. While 
          the first one develops mostly in the inner regions of the star, 
          the second one is favored in the outer regions, where the 
          Rossby number is much smaller and a mean-field dynamo action is 
          more efficient. By solving the mean-field induction equation we 
          have computed the critical spin period below which no dynamo 
          action is possible and found it to be $\sim 1$ s for a wide 
          range of stellar models and for both axisymmetric and 
          non-axisymmetric magnetic fields. Because this critical period 
          is substantially longer than the characteristic spin period of 
          very young pulsars, we expect that a mean-field dynamo will be 
          effective for most protoneutron stars. The saturation dipole
          field estimated by making use of the model of ``global'' 
          quenching fits well the pulsar magnetic fields inferred from
          the spin-down data. Apart from the large scale magnetic field,
          our model predicts also a generation of small scale fields
          which are typically stronger than the poloidal field and can
          survive during the lifetime of pulsars. Extremely rapidly rotating
          protoneutron stars ($P \sim 1$ ms) may have the dipole field $\sim
          (3-6) \times 10^{14}$ G.   
 }

\titlerunning{Mean-field dynamo action in PNSs}

\maketitle

\section{Introduction}
The origin of pulsar magnetism is a subject of debate for decades.
In a simple magnetic dipole braking model, the polar field strength inferred
from observational data can reach $\sim 5 \times 10^{13}$G. One possibility 
is that the magnetic field of a progenitor star is amplified by many orders 
of magnitude because of the conservation of the magnetic flux during the
collapse stage (Ginzburg 1964, Woltier 1964). However, the ``fossil field'' 
hypothesis despite being seemingly attractive and plausible meets a number 
of problems. For instance, the progenitor star should possess a sufficiently
strong magnetic field that does not agree with observational data (see 
discussion in Thompson \& Duncan 1993 for more details).

Another possibility is a turbulent dynamo action that can amplify the
magnetic field during first $\sim 30-40$s of a neutron star life when the 
star is the subject of hydrodynamic instabilities. Hydrodynamic 
instabilities in protoneutron stars (PNSs) are driven by either lepton 
gradients which result in the so-called ``neutron-finger instability'' 
(Bruenn \& Dineva 1996), or by negative entropy gradients which are common 
in simulations of supernovae (Bruenn \& Mezzacappa 1994, Rampp \& Janka 
2000) and in models of PNSs (Keil \& Janka 1995; Keil, Janka \& M\"{u}ller 
1996; Pons et al. 1999). The latter instability is usually referred to as 
the ``convective instability''. The nature of instabilities in PNSs has 
been considered by a number of authors (Grossman, Narayan \& Arnett 1993, 
Bruenn \& Dineva 1996, Miralles, Pons \& Urpin 2000). Turbulent motions 
caused by instabilities in combination with rapid rotation that seems to 
be almost unavoidable in PNSs make turbulent dynamo one of the most 
plausible mechanism responsible for the pulsar magnetism.

Recent measurements of X-ray spectra of some pulsars have started to 
provide a closer look at the magnetic field at the neutron star surface.
These observations indicate that the pulsar magnetic field may often 
have a fine structure near the surface apart from the global magnetic 
structure inferred from the spin-down data. For instance, the absorption 
features in the spectrum of 1E 1207.4-5209 allow to estimate a surface 
magnetic field as $\sim 1.5 \times 10^{14}$G (Sanwal et al. 2002), that 
is in contrast with the dipolar magnetic field estimated from the spin-down 
rate of this pulsar ($\sim (2-4) \times 10^{12}$G (Pavlov et al. 2002)). 
Becker et al.(2003) have found an emission line in the X-ray spectrum of 
PSR B1821-24 that could be interpreted as cyclotron emission from the 
pulsar's polar cap. The line can be formed in a magnetic field $\sim 3 
\times 10^{11}$G, approximately two orders of magnitude stronger than the 
dipolar magnetic field. Kargaltsev, Pavlov \& Romani (2004) have reported 
a marginal detection of the emission line in the spectrum of the 
millisecond pulsar J0437-4715 that can be interpreted as a cyclotron line 
from the magnetic sport with the field $\sim 7 \times 10^{8}$G on the 
neutron star surface. Haberl et al.(2003) found evidence for a proton 
cyclotron line in the spectrum of the isolated pulsar RBS 1223. This line 
corresponds to the local magnetic field at the stellar surface 
of approximately one order of magnitude stronger than the dipole field. All these 
measurements provide evidence that the magnetic field likely has a complex 
structure in neutron stars with small scale fields being stronger than (or
comparable to) the large scale field responsible for the secular spin-down 
of pulsars.

Observations of radio emitting pulsars also support the idea that neutron 
stars apart from the large scale field (likely, dipoles) may have small scale 
magnetic structures at the surface. Recently, Gil \& Mitra (2001) and Gil \& 
Melikidze (2002) have argued that the formation of a vacuum gap in radio 
pulsars is possible if the magnetic field lines near the polar cap have a 
small curvature $\sim 10^{5}$cm and the field is very strong, $B_{s} \sim 
10^{13}$G, irrespective of the magnetic field measured from the spin evolution.
Furthermore, the presence of a strong magnetic field with a small curvature 
can account for the radio emission of many radiopulsars that lie in the pulsar 
graveyard and should be radio silent (Gil \& Mitra 2001). Analysing drifting 
subpulses observed in many pulsars, Gil \& Sendyk (2003) found that their 
behavior is consistent with the vacuum gap maintained by a strong spot-like 
magnetic field.

The growing number of evidences for a complex structure of the magnetic field 
suggests that this may represent a general phenomenon in neutron stars. The 
presence of a large scale field accompanied by small scale magnetic structures
can naturally be understood if the field is generated by the turbulent 
mean-field dynamo. The duration of an unstable stage in PNSs ($\sim 30-40$ s) 
is sufficient for dynamo to reach a saturation level because the period of 
rotation of PNSs and the dynamo growth time are typically much shorter 
(Thompson \& Duncan 1993). The magnetic fields generated by turbulent dynamo 
in convective PNSs will be frozen in the crust that starts to form almost 
immediately after convection stops. Since the crustal conductivity is high 
both the large and relatively small scale fields can survive in neutron 
stars during a long time comparable to the lifetime of pulsars (Urpin \& 
Gil 2004). 

Recently, it has been shown by Bonanno, Rezzolla \& Urpin (2003) that 
turbulence can drive both small and large scale dynamos in PNSs. This 
result is in contrast to the previous conclusion by Thompson \& Duncan (1993)
that only small scale dynamo can be operative in neutron stars. The
reasoning of these authors was based on the assumption that the whole PNS
is convectively unstable with turbulent velocity $v_{T} \sim 10^{8}$ 
cm s$^{-1}$. An efficiency of the mean-field dynamo can be characterized
by the Rossby number, $Ro= P/\tau_{T}$, where $P$ is the period of rotation
and $\tau_{T}$ is the turnover time of turbulence. If $P \sim 10-100$ ms
that is likely typical for young neutron stars (Narayan 1987) and $\tau_{T} 
\sim 1$ ms that is typical for the convectively unstable region of PNSs 
then $Ro \sim 10-100$ and the influence of rotation on the turbulence is 
therefore weak. As a consequence, Duncan \& Thompson (1992) and Thompson \& 
Duncan (1993) concluded that the mean-field dynamo does not operate in PNSs 
except those rotating with the period $\sim 1$ ms. 

The model of pulsar magnetism qualitatively similar to that by 
Thompson \& Duncan (1993) has been outlined by Wheeler et al. (2000) 
and Wheeler, Meier \& Wilson (2002) who pointed out the importance
of differential rotation that seems to be almost unavoidable in PNSs.
Their model suggests that a very strong toroidal magnetic field can 
be generated by differential rotation in a newly born pulsar. These 
authors assume that a poloidal field of $\sim 10^{12}$ G could arise 
simply from flux-freezing if the precollapse progenitor core has a 
field strength comparable to that of a magnetized white dwarf, $\sim 
10^{8}$ G. Mention that this point seems to be in contradiction with 
observational data and has been criticized by Thompson \& Duncan 
(1993). The poloidal field can be amplified further by differential 
rotation to produce a strong toroidal field. The toroidal field is 
growing linearly with time, and the growth is limited by buoyancy 
which operates to expel the field from the site where it is generated. 
According to the authors, the toroidal field at the buoyancy limit 
can be as strong as $\sim 10^{16}$ G. Generally, such a strong field 
can be sufficient to suppress convection in proto-neutron stars 
(Miralles, Pons \& Urpin 2002). Note, however, that the turbulent 
magnetic diffusion caused by convection can essentially decrease the 
saturation toroidal field as it will be seen from our calculations and,
most likely, the generated field will not suppress convective motions.

In fact, the picture is more complex since two essentially different 
instabilities may occur in PNSs (Bruenn \& Dineva 1996, Miralles, Pons
\& Urpin 2000). Convection is presumably connected to the entropy 
gradient, whereas the neutron-finger instability is more relevant to a 
negative lepton gradient. The neutron-finger instability is the 
astrophysical analogy of the salt fingers that exist in terrestrial
oceans. Physically, a fluid parcel perturbed downward in the PNS can
thermally equilibrate more rapidly with the background but find itself
lepton-poorer and denser and, therefore, subject to a downward force
that would amplify perturbations. The neutron-finger instability is
a sort of the so called ``doubly diffusive'' instabilities that occur
due to dissipative processes like viscosity, thermal and lepton 
diffusivity. These processes are rather fast in PNSs and, therefore, 
the secular neutron-finger instability can grow rapidly. The 
estimated growth time in the neutron-finger unstable region, $\sim
30-100$ ms, is only a couple orders of magnitude longer than the growth 
time of extremely rapid convection (Miralles, Pons, \& Urpin 2000), 
thus yielding a mean turbulent velocity $\sim (1-3) \times 10^{6}$ cm/s.

Typically, the convectively unstable region is surrounded by the 
neutron-finger unstable region, the latter involving therefore a larger 
portion of the stellar material. As a result, not the whole PNS is 
convectively unstable in contrast to the assumption made by Thompson \&
Duncan (1993). Instabilities first develop in the outer layers containing
$\sim 30\%$ of the stellar mass, The two unstable 
regions move towards the inner parts of the star and, after $\sim 10$ s, 
more than $90\%$ of the stellar mass is hydrodynamically unstable. At this 
stage the stellar core has become convectively unstable but it is still 
surrounded by an extended neutron-finger unstable region. In $\sim 20 $ 
s that follow, the temperature and lepton gradients are progressively 
reduced and the two unstable regions begin to shrink, leaving the outer 
regions of the star. After $\sim 30$ s, most of the PNS is stable and 
the instabilities disappear completely after $\sim 40$ s (Miralles, Pons 
\& Urpin, 2000). The Rossby number is actually large in the convectively
unstable region, $Ro \sim 10-100$, and the mean-field dynamo likely
does not work in this region. However, this is not the case in the neutron
finger unstable region where the turbulent velocity is much slower than
in the convectively unstable region, and $Ro \sim 1$ (Bonanno, Rezzolla
\& Urpin, 2003). Therefore, turbulence can be strongly modified by
rotation in the neutron finger unstable region, and this favors the
efficiency of a mean-field dynamo. Of course, in both regions turbulent
magnetic fields can also be generated by small scale dynamo.

In the present paper, we consider the mean-field dynamo action in PNSs
in more detail. We extend our study to the case of non-axisymmetric
fields that are of particular interest in pulsars. The main goal of this
paper is to show that the mean-field dynamo can operate in a wide range
of the parameters of PNSs and generate the field of the strength
comparable to that of ``standard'' pulsars. The paper is organized as 
follows. In Section 2, we consider the basic equations governing the 
mean-field dynamo and discuss the properties of convection in PNSs. The 
numerical results are represented in Section 3, and our findings are 
summarized in Section 4.

\section{Basic equations}

    To investigate the efficiency of a mean-field dynamo action, we model 
the PNS as a sphere of radius $R$ with two substantially different turbulent 
zones separated at $R_{c}<R$. The inner part ($r<R_{c}$) corresponds to the
convectively unstable region, while the outer one ($R_{c}<r<R$) to the
neutron-finger unstable region. The boundary between the two regions moves 
inward on a timescale comparable to the cooling timescale (i.e. $\sim 1-10$ 
s) that is much longer than the turnover time for the both unstable zones. 
The energy of turbulence is generally non-stationary as well, developing 
rapidly soon after the collapse, reaching a quasi-stationary regime after a 
few seconds, and then progressively disappearing. However, changes take place
on the cooling time scale and, therefore, all parameters of turbulence can be 
treated in quasi-steady approximation. In this case, the mean-field induction
equation for a turbulent PNS can be written as
\begin{equation}
\frac{\partial \vec{B}}{\partial t} = \nabla \times (\vec{v} 
	\times \vec{B} + \alpha \vec{B})
	- \nabla \times (\eta \nabla \times \vec{B}) \ ,
\end{equation}
where $\eta$ is the turbulent magnetic diffusivity, $\alpha$ is a pseudo-scalar 
measuring the efficiency of the dynamo (the ``$\alpha$-parameter''), and 
$\vec{v}$ is the velocity of a mean fluid motion. Boundary conditions for 
the magnetic field need to be specified at the stellar surface, where we impose 
vacuum boundary conditions, and at the center of the star, where we impose the 
vanishing of the toroidal magnetic field.

We assume that differential rotation is the only large scale motion
in PNSs, and $\vec{v} = \vec{\Omega}(\vec{r}) \times \vec{r}$. From 
theoretical modeling and simple analytic considerations it is commonly 
accepted that core collapse of a rotating progenitor leads to differential
rotation of a newly born neutron star (Zwerger \& M\"{u}ller 1997; Rampp, 
M\"{u}ller \& Ruffert 1998; Dimmelmeier, Font \& M\"uller 2002) mainly due
to conservation of the angular momentum during collapse. A recent study on
evolutionary sequences of rotating PNSs (Villain et al. 2004) shows that the
typical scale on which the angular velocity changes is in the range $\approx
5-10$ km. We consider two possible models of differential rotation in 
PNSs, a shellular rotation with
\begin{equation}
\Omega (r) = \Omega_{sph}^{(0)} + \left( \frac{r}{R} \right)^{2} 
\Omega_{sph}^{(1)} \ ,
\end{equation}
and cylindrical rotation with
\begin{equation}
\Omega (s) = \Omega_{cyl}^{(0)} + \left( \frac{s}{R} \right)^{2} 
\Omega_{cyl}^{(1)} \ ,
\end{equation}
where $r$ and $s$ are the spherical and cylindrical radii, respectively.
{ Note that $\Omega(\vec{r})$ can generally depend on the time 
because the PNS cools down and characteristics of turbulence change
during the unstable stage. Turbulent transport of the angular
momentum takes place on the diffusive timescale, $\sim R^{2}/
v_{T}^{2} \tau_{T}$, that is short compared to the cooling time of
PNSs. Therefore, differential rotation is likely constrained to 
the properties of turbulent motions at each instant of time, and
$\Omega(\vec{r})$ follows adiabatically the PNS thermal and
chemical evolution. We will neglect these relatively slow changes 
in our quasi-steady modeling. This can well be justified because 
the main conclusions of our study are qualitatively the same for 
any considered rotation law including rigid rotation.} 

As it was mentioned, all of the PNS undergoes turbulent motions but with 
properties that are different in the inner parts ($0 \leq r \leq R_{c}$), where 
fast convection operates, from those in the outer parts ($R_{c} \leq r \leq R$), 
where the neutron-finger instability operates. To model this in a simple way, we 
assume the relevant physical properties of the two regions to vary in a smooth 
way mostly across a thin layer of thickness $\Delta R = 0.025 R$. More
precisely, we express $\eta$ as
\begin{equation}
\eta = \eta_c + (\eta_{nf}-\eta_c) \left\{1+
	erf[(r-R_c)/\Delta R] \right\}/2 \ ,
\end{equation}
where $\eta_{c}$ and $\eta_{nf}$ are respectively the turbulent magnetic
diffusivities in the convective and neutron-finger unstable zones, 
and {\it erf} is the ``error function''. Using equation (4), we have $\eta 
\approx \eta_{c}$ if $R_{c} - r \gg \Delta R$ (convective zone), while $\eta 
\approx \eta_{nf}$ if $r-R_{c} \gg \Delta R$ (neutron finger unstable zone). 
In the both unstable zones, the turbulent magnetic diffusivity can be
estimated as $v_{T} \ell_{T}/3$ where $v_{T}$ and $\ell_{T}$ are the
corresponding turbulent velocity and length-scale. The turbulent
length-scales are likely comparable in the both zones and $\sim 1-3$ km.
The turbulent velocities are, however, different with much larger velocity
in the convective zone (see Bonanno, Rezzolla \& Urpin 2003). Therefore,
$\eta_{nf} \ll \eta_{c}$ in PNSs, and we choose $\eta_{nf}/\eta_c = 0.1$
in calculations. 

Similarly, we have modeled the $\alpha$-parameter as being negligibly 
small in the convectively unstable region and equal to $\alpha_{nf}$ in 
the neutron-finger unstable region, i.e.
\begin{equation}
\label{alp}
\alpha(r,\theta) = \alpha_{nf}\cos\theta \left\{1 + 
	erf[(r-R_c)/{\Delta R}] \right\} /2 \ ,
\end{equation}
where the angular dependence chosen in this expression is the simplest
guaranteeing antisymmetry across the equator. Usually, we assume 
$\alpha_{nf} =$const in calculations. However, the $\alpha_{nf}$-parameter 
is likely not constant in the neutron finger unstable region because
differential rotation or inhomogeneity of turbulence can generally produce 
some non-uniformity in $\alpha$. For instance, a non-uniformity can originate
from overshooting caused by convection that smooths the boundary layer 
between the zones. To understand how much our conclusions are sensitive 
to the choice of $\alpha$, we performed some calculations with a radially
dependent $\alpha_{nf}(r)=\alpha_{nf0} F(r)$ where $F(r)$ is plotted in 
Fig.1.

\begin{figure}
\begin{center}
\includegraphics[width=9.0cm]{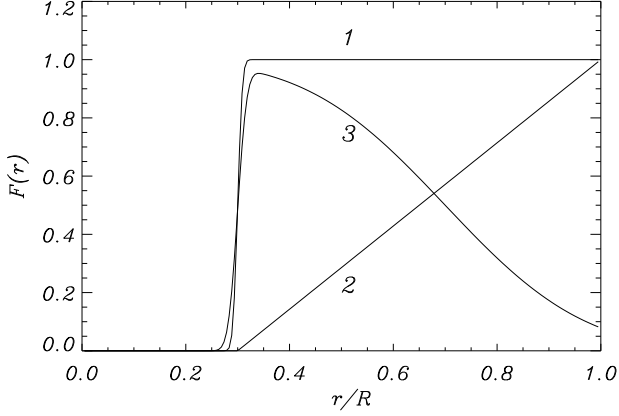}
\caption{The sketch of the radial dependence of $F(r)$ for the three models
used in our calculations.}
\end{center}
\end{figure}

We recall that in a rotating turbulence with length-scale $\ell_{T}$
and moderate Rossby number, $\alpha_{nf0} \approx \Omega \ell_{T}^{2} \nabla \ln
(\rho v_{_{T}}^{2})$ (R\"{u}diger \& Kitchatinov 1993). 
In PNSs, however,
the pressure is mainly determined by number density of degenerate neutrons
and, therefore, the density length-scale
is comparable to the pressure one $L$ and to
the maximal length-scale of the instabilities. This introduces a great
simplification since we have then $\alpha_{nf0} \approx \Omega L$.

\section{Numerical results}
The induction equation (1) with $\Omega$, $\eta$ and $\alpha$ given
by equations (2)-(3), (4) and (5) has been solved with a numerical code
employing finite-difference techniques for the radial dependence and a
polynomial expansion for the angular dependence. In particular, we use 
the representation (R\"{a}dler, 1973)
\begin{equation}
\vec{B}= \vec{B}_{\rm p} + \vec{B}_{\rm t}, \;\;\;\;\;\;\; 
\vec{B}_{\rm p}= \nabla \times \vec{A}_{\rm t}
\end{equation}
with 
\begin{equation}
\vec{B}_{\rm t} = -\vec{r}\times\nabla \Psi, \;\;\;\;\;\;\; \vec{A}_{\rm t} = - \vec{r} \times \nabla \Phi
\end{equation}
and the scalar functions $\Psi$ and $\Phi$ have the following expansion
\begin{eqnarray}
&&\Phi = \sum_{n=1}^{\infty} \sum_{m=-n}^{n} e^{-i\lambda_{mn}t} \Phi_{n m}(r) Y_n^m(\theta,\phi)\nonumber\\[2mm]
&&\Psi = \sum_{n=1}^{\infty} \sum_{m=-n}^{n}  e^{-i\lambda_{mn}t}\Psi_{n m}(r) Y_n^m(\theta,\phi)
\end{eqnarray}
The induction equation thus decouple in a $m \times n$-dependent set of coupled 
equations for $\Psi$ and $\Phi$ which can be solved by truncating the order 
$n$ of the harmonics for a given value of $m$ (further details can be found
in R\"{a}dler, 1973, Bonanno {\it et al}, 2002). 
Non axisymmetric modes, {\it i.e.} those with $m\not =0$  are waves traveling 
in azimuthal direction. The field configuration of a non-axisymmetric mode 
rotates like a rigid body with 
angular velocity $\lambda_{mn}/m$.
The simulations reported 
here use 30 spherical harmonics and about 40 grid points in the radial 
direction in order to ensure convergence. Boundary conditions are chosen 
in order to guarantee the regularity at the center and vacuum outside $r=R$. 
We have solved the induction equation (1) to determine the critical
value $\alpha_{0}$ corresponding to the marginal stability of the
dynamo. The seed magnetic field will grow if $\alpha_{nf0} >
\alpha_{0}$ and decay if $\alpha_{nf0} < \alpha_{0}$. Since
$\alpha_{nf0} \approx \Omega L$, the critical value $\alpha_{0}$
effectively selects a critical value for the spin period, $P_{_{0}}
\equiv 2 \pi L/ \alpha_{0}$, such that magnetic field generation via
a mean-field dynamo action will be possible only if the stellar spin
period is shorter than the critical one. The different types of dynamo
can be distinguished according to whether the generated field exhibits
periodic oscillations (oscillatory dynamo, dashed lines in Fig.2) or
not (stationary dynamo, solid lines)

In Fig.2, we plot the critical period $P_{0}$ 
for the case of the shellular rotation (2). The
critical period is shown as a function of the parameter $q_{sph}=
\Omega_{sph}^{(1)}/(\Omega_{sph}^{(0)}+\Omega_{sph}^{(1)})$ that
characterizes differential rotation. Note that $q > 0$ and $q < 0$ 
correspond to situations in which the stellar surface rotates faster 
and slower than the center, respectively. Calculations have been done 
for $\eta_{nf} = 10^{11}$ cm$^2$ s$^{-1}$ and $F(r) =1$ (constant 
$\alpha$-parameter in the neutron finger unstable region).
The different types of dynamo 
can be distinguished according to whether the generated field exhibits 
periodic oscillations (oscillatory dynamo, dashed lines in Fig.2) or 
not (stationary dynamo, solid lines). 
\begin{figure}
\begin{center}
\includegraphics[width=9.0cm]{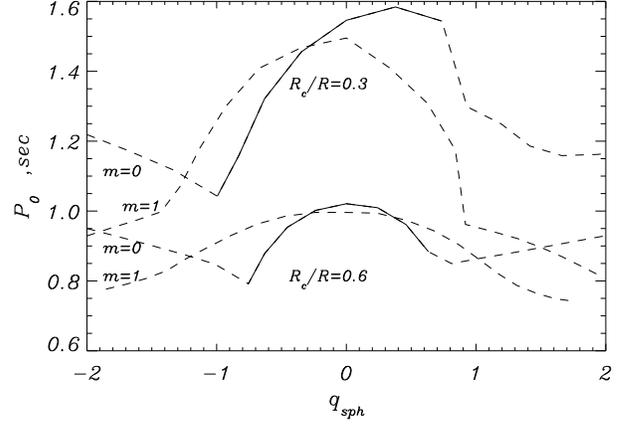}
\caption{Critical period as a function of the differential rotation
parameter $q_{sph}$ for $F(r)=1$. The two pairs of curves refer to 
different values of $R_{c}$, with the solid parts corresponding to a 
stationary dynamo and the dashed parts to an oscillatory dynamo. The 
number $m$ corresponds to different values of the azimuthal wavenumber 
in a polynomial expansion for the magnetic field. }
\end{center}
\end{figure}

As shown in Fig.2, a stationary dynamo dominates the axisymmetric 
magnetic field generation process for $|q| \lesssim 1$, while an oscillatory
dynamo is more efficient for $1 \lesssim |q|$. These two regimes correspond 
to $\alpha^2$-dynamo and $\alpha\Omega$-dynamo, respectively. Given the 
large differential rotation required for a generation of the axisymmetric 
oscillatory magnetic field, it may be difficult to achieve in practice. Hence, 
the $\alpha^{2}$-dynamo appears to be the most likely source of axisymmetric 
magnetic field generation via dynamo processes in PNSs. The situation is, 
however, different for the generation of a non-axisymmetric magnetic field 
that is of interest for PNSs since the observed pulsars have non-axisymmetric 
fields. 


The critical spin period found in our calculations is in general rather long. 
For instance, a mean-field dynamo will develop if $P \leq 1$ s when $R_{c}/R 
=0.6$ and if $P \leq 1.4$ s when $R_{c}/R=0.3$ if a PNS rotates rigidly (i.e. 
$q_{sph}=0$). If the PNS rotates differentially, the critical period is 
typically reduced. The difference, however, is not large except the case 
of the non-axisymmetric field ($m=1$) in a star with $|q_{sph}| >1$ when the 
difference can reach $\sim 40$\%. As a result, only PNSs with $P \gtrsim 1$ 
s will not be subject to a turbulent mean-field dynamo action. Such slow 
rotation rates should be rather difficult to achieve if angular momentum is 
conserved during the collapse to a PNS. We expect, therefore, that a 
turbulent mean-field dynamo will be effective during the initial stages of 
the life of most PNSs.

In Fig.(3) and (4), we plot the same as in Fig.(2) but for $F(r)$ represented by 
the lines 2 and 3 in Fig.(1), respectively. The behavior of $P_{0}$ is 
qualitatively unchanged
and the only difference is that there is no 
stationary dynamo regime for a generation of the axisymmetric field in the 
case $R_c/R=0.6$ if the function $F(r)$ is decreasing to the surface (the model
3 in Fig.(1)). Like the case $F(r)=1$, the critical period is longer for 
$R_c/R=0.3$ compared to $R_c/R=0.6$. The difference is due to the fact that 
a PNS with a more extended neutron-finger unstable region can rotate 
proportionally more slowly while maintaining the same dynamo action. The 
values of $P_{0}$ are typically by a factor $\sim 2$ smaller than for a 
dynamo action with $F(r)=1$ implying that the onset of dynamo is mainly 
characterized by the average value of $\alpha$ in the unstable region rather 
than a particular dependence of $\alpha$ on $r$.  

\begin{figure}
\begin{center}
\includegraphics[width=9.0cm]{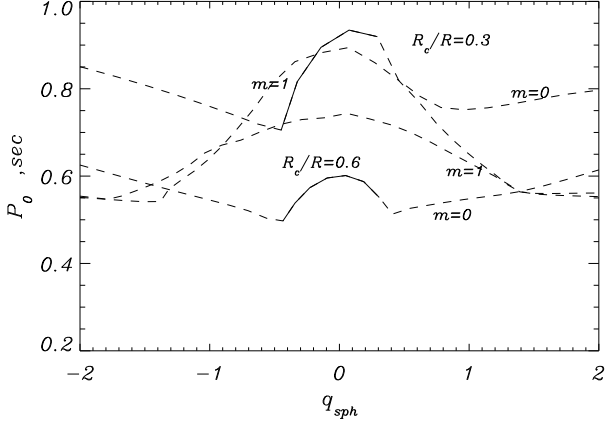}
\caption{Critical period as a function of the differential rotation
parameter $q_{sph}$ for $F(r)$ given by the model 2 in Fig.1. 
Other parameters are the same as in Fig.2.}
\end{center}
\end{figure}

\begin{figure}
\begin{center}
\includegraphics[width=9.0cm]{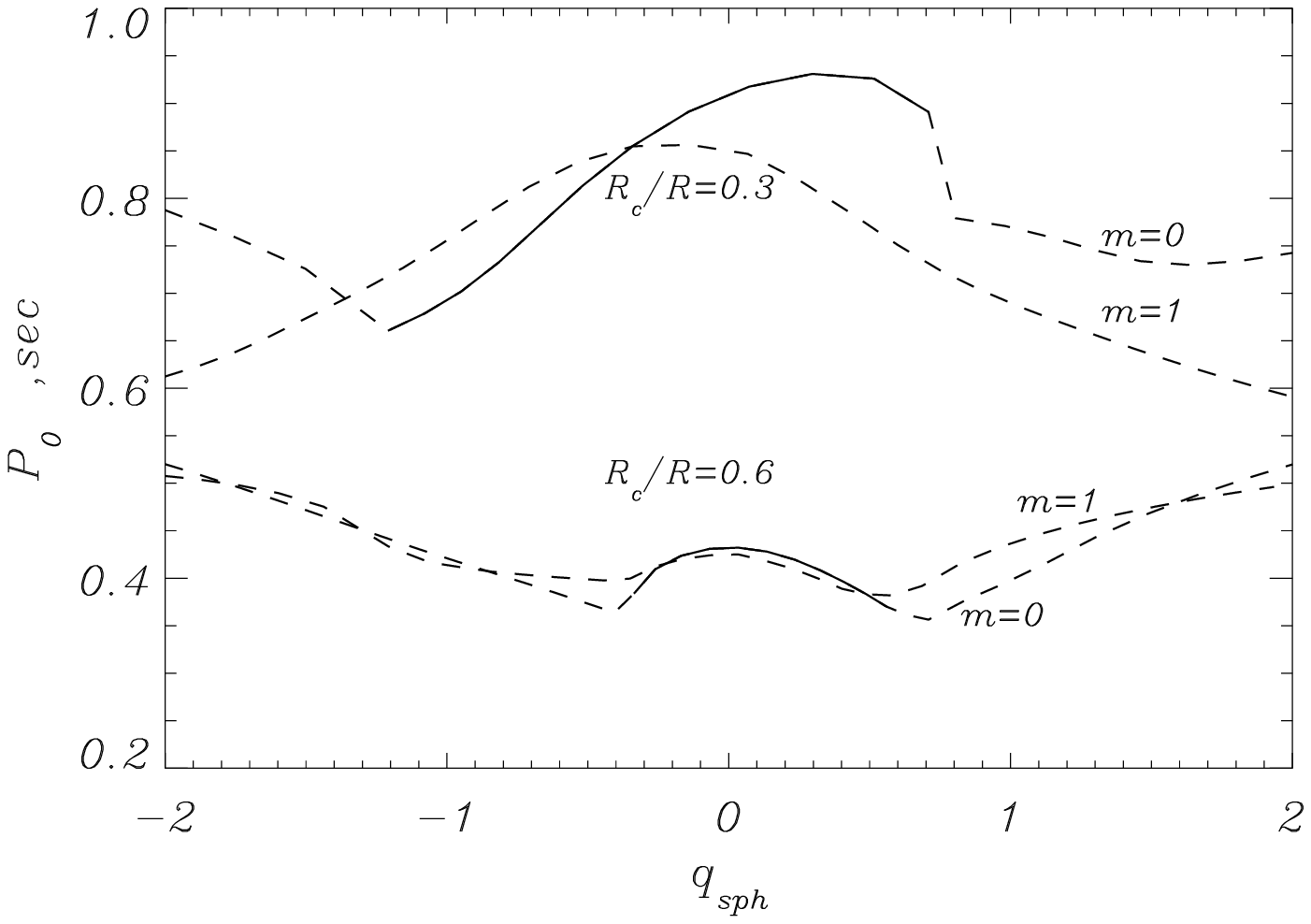}
\caption{Critical period as a function of the differential rotation
parameter $q_{sph}$ for $F(r)$ given by the model 3 in Fig.1. 
Other parameters are the same as in Figs.2 and 3.}
\end{center}
\end{figure}

\begin{figure}
\begin{center}
\includegraphics[width=9.0cm]{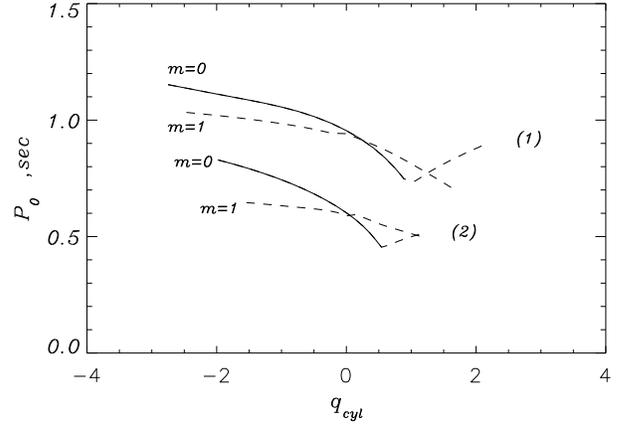}
\caption{Critical period as a function of the differential rotation
parameter $q_{cyl}$ for $F(r)$ given by the models $(1)$ and $(2)$ in Fig.1.
for $R_c/R=0.6$ . 
\label{fig5}}
\end{center}
\end{figure}

\begin{figure}
\includegraphics[width=9.0cm]{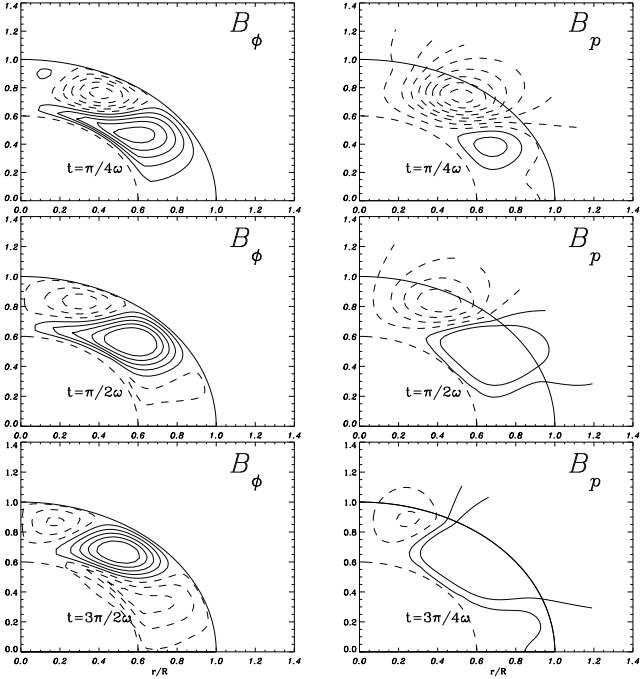}
\caption{Toroidal ($B_{\phi}$) and the poloidal ($B_p$) magnetic field
lines for different pulsational times for $m=0$ and $q\approx 2$ 
in Fig.(\ref{fig5}). Solid and dashed contours correspond to positive 
and negative values, respectively. The dot-dashed line marks the 
position of $R_c = 0.6 R$. \label{mag}}
\end{figure}

The above discussion for the shellular model Eq.(2), remains 
qualitatively similar for a cylindrical rotational law Eq.(3) 
for $q_{cyl}= \Omega_{cyl}^{(1)}/(\Omega_{cyl}^{(0)} + 
\Omega_{cyl}^{(1)})$. Since in this case the  differential rotation is not 
important  the basic dynamo action is that of an 
$\alpha^2$-model. If instead $q_{cyl} \gtrsim 1$, the dynamo mechanism 
would produce an oscillating field of an $\alpha\Omega$-type. The 
general dependence of the critical period on the differential rotation 
parameter $q_{cyl}$ is shown in Fig.(\ref{fig5}) for $R_c = 0.6 R$. 
Note that, for small $q_{cyl}$, the $m=1$ modes have a greater critical 
periods than the $m=0$ modes, and thus the critical value of $\alpha_{nf0}$
for non-axisymmetric modes is smaller than for the axisymmetric case. As 
a consequence, non-axisymmetric field is more easily excited than the 
axisymmetric one in this case. 

In Fig.(\ref{mag}), we show the toroidal ($B_{\phi}$) and poloidal
($B_{p}$) magnetic fields of a typical PNS model for the cylindrical 
rotation law. Note that the both components are generated in the outer 
neutron-finger unstable region but turbulent diffusion transports the
magnetic field also in the inner convectively unstable region. This 
field is, however, considerably weaker. Our calculations show that 
if $|q|<1$ and the field is generated by the $\alpha^{2}$-dynamo then 
$B_{\phi}/B_{p} \sim 10$, while $B_{\phi}/B_{p} \sim 100-200$ if 
$|q|>1$ and the $\alpha \Omega$-dynamo generates the magnetic field. 
Both results suggest that the internal magnetic fields in neutron 
stars could be substantially stronger than the observable surface 
fields. The toroidal magnetic field tends to concentrate near the 
polar regions whereas the poloidal one is more evenly distributed 
in latitude. Note that, in the case of an $\alpha^2$-dynamo, the 
generated field propagates more efficiently into the convective zone 
(see Bonanno, Rezzolla and Urpin 2003), 
\section{Discussion}
The PNS is subject to two substantially different instabilities, with a 
convective instability operating in the inner region of the star and a 
neutron-finger instability being more efficient in the outer region. The 
turbulent motions are more rapid in the convective zone, where the Rossby 
number is large but the $\alpha$-parameter, characterizing the mean-field 
dynamo action, is likely small. In the neutron-finger unstable region, on 
the other hand, the turbulent turnover time is considerably longer, the 
Rossby number small, and the $\alpha$-parameter can be sufficiently large 
to drive a mean-field dynamo. 

The $\alpha$-parameter depends the stellar rotation being larger for 
rapidly rotating stars. Our simulations show that even relatively slowly
rotating PNSs can be subject to a dynamo action, with the $\alpha^2$-dynamo 
being the most efficient mechanism of generation for both axisymmetric and
non-axisymmetric fields if differential rotation is not extremely strong. 
The calculated critical value of the spin period that determines the onset 
of dynamo in PNSs is $P_{_{0}} \sim 1 $ s for a wide range of models. This 
value is essentially larger even than the characteristic spin period of 
very young pulsars as inferred from observations ($\sim 50-100$ ms, Narayan 
1987) but likely PNSs can rotate even faster. As a result, a turbulent
mean-field dynamo can be effective in the early stages of the life of most 
PNSs. The critical period is not very different for axisymmetric and 
non-axisymmetric fields and likely both these magnetic configurations can 
be generated by the PNS dynamo. The generation of a non-axisymmetric 
component is the key point for the magnetism of pulsars since they have a 
substantial non-axisymmetric field. It must be stressed that, as the 
neutron-finger unstable region shrinks towards the surface, the difference 
between axisymmetric  and non-axisymmetric critical periods, disappears 
as it can be noticed, for instance, in  Fig.~2 and Fig.~3 thus providing 
an opportunity to generate a non-axysimmetric field in a very general 
framework.

The PNS dynamo can operate both in oscillatory and stationary regime.
A non-axisymmetric field is always a wave traveling in the azimuthal direction.
On the other hand, axisymmetric field can develop in both the oscillatory and
stationary regimes, depending on the profile of $\alpha$ in the
neutron finger unstable region and the thickness of this region. The 
period of dynamo oscillations can be estimated as $\sim 0.1 
R^{2}/ \eta_{nf} \sim R^{2}/\eta_{c}$ and, in our model, is $\sim 1$ s. 
Note that, for the pulsar magnetism, there is no particular difference
in which regime, oscillatory or stationary, the field is generated because
a formation of the crust starts almost immediately after the instability
stops, and the generated field should be frozen-in a highly conductive 
crustal matter. 

The unstable stage lasts $\sim 30-40$ s in PNSs, and likely this time is
sufficient for the dynamo to reach a saturation level. We estimate the 
field strength in saturation by making use of the model of a ``global'' 
quenching. In accordance with this model, a generation of the magnetic 
field decreases the $\alpha$-parameter, and the simplest possible
expression can be proposed for the non-linear $\alpha$-parameter in the 
neutron finger unstable zone,
\begin{equation}
\alpha_{nl}(\tilde{B}) = \frac{\alpha_{nf0}}{1 + \tilde{B}^{2}/B_{eq}^{2}},
\end{equation}
where $\tilde{B}$ is the characteristic value of the generated field,
and $B_{eq}$ is the equipartition magnetic field determined by equating 
the kinetic and magnetic energy of turbulence. Then, the saturation 
magnetic field, $B_{sat}$, can be estimated if we assume that 
$\alpha_{nl}(B_{sat})$ is equal to the critical value $\alpha_{0}$ that 
determines the marginal dynamo stability. We have from this condition
\begin{equation}
B_{sat} \approx B_{eq} \sqrt{\frac{P_0}{P}-1},
\end{equation}
If $P_{0} > P > P_{0}/2$, the generated mean field is weaker than the
turbulent magnetic field, $B_{eq}$. On the contrary, the saturation  
field is stronger than  $B_{eq}$ if $P < P_{0}/2$, and $B_{sat}$ can 
reach a rather large value if the PNS rotates very rapidly. Likely, 
however, that the rotation period cannot be shorter than $\sim 1$ ms 
and, hence, the maximum field generated by dynamo in PNSs is around 
$\sim 30 B_{eq}$. In our calculations, the generated toroidal field 
is typically stronger than the poloidal field. Therefore, quenching 
is determined mostly by the strength of the toroidal field, and 
equation (10) provides an estimate of the saturation toroidal field. 
Since, in the case of an $\alpha^{2}$-dynamo, the poloidal field is 
approximately 5-10 times weaker, we obtain for the saturation poloidal 
field 
\begin{equation}
B_{p \; sat} \approx (0.1-0.2) B_{eq} \sqrt{\frac{P_0}{P}-1}.
\end{equation}
The poloidal field is typically weaker than (or comparable to) the
equipartition field. The only exception is very rapidly rotating PNSs 
with the period $\sim 1-3$ ms where the poloidal field can be a factor 
few stronger than small-scale turbulent fields. In the majority of 
pulsars, however, we can expect that a large scale poloidal (for 
example, dipolar) field should not be stronger than the small-scale 
fields if their magnetism is caused by the mean-field dynamo action. 
Note that because of a high conductivity, turbulent magnetic fields 
with the length-scale $\gtrsim 1$ km can survive in the neutron star 
crust for a very long time comparable to the lifetime of pulsars, $\sim 
100$ Myr (Urpin \& Gil 2004).  
   
Turbulence is non-stationary in the both unstable zones of PNSs, and 
therefore $B_{eq}$ in equation (9)-(11) varies with time. It rises 
very rapidly soon after collapse, reaches some quasi-steady regime, 
and then goes down when the temperature and lepton gradients are 
smoothed (after $\sim 30-40$ s). The timescale required for fluid to 
make one turn in a turbulent cell 
with the length-scale $\ell_{T}$ can be estimated as $\tau_{T} \approx \pi 
\ell_{T} / v_{T}$. This timescale varies with time as well, but is 
typically much shorter than the characteristic cooling timescale of the
PNS, $\tau_{cool}$, except the very late phase when gradients are smoothed 
and instabilities are less efficient. Therefore, turbulence can be treated 
in a quasi-steady approximation during almost the whole convective 
evolution of the PNS except the late stage. We can estimate $B_{eq}$ during 
the quasi-steady regime as $\sim 10^{16}$ G in the convective zone, and 
$\sim (1-3) \times 10^{14}$ G in the neutron finger unstable zone (see Urpin 
\& Gil 2004). However, the temperature and lepton number gradients are
progressively reduced as the PNS star cools down and, therefore, 
the turbulent velocity decreases as well. As a result, the strength of 
small-scale magnetic fields generated by turbulence also decreases compared 
to the maximum value, but the turnover time of turbulence increases. The 
mean-field dynamo as well as the small-scale one are still operative until 
the quasi-steady condition $\tau_{cool} \gg \tau_{T}$, is fulfilled. We 
assume that this condition breaks down at some instant of time when 
$\tau_{T}$ becomes comparable to the cooling timescale: $\tau_{T} \sim 
\tau_{cool}$. Then, the turbulent velocity at this instant is given by
\begin{equation}
v_{T} \sim \frac{\pi \ell_{T}}{\tau_{cool}}.
\end{equation}
We assume that the final strength of the magnetic field generated
by both the mean-field and small-scale dynamo is determined by 
$B_{eq}$ at the instant of time when the quasi-steady condition 
breaks down. 
Therefore, we have for the final strength of the equipartition
magnetic field
\begin{equation}
B_{eq} \sim  \sqrt{4 \pi \rho} v_{T} \sim 
\frac{\pi \sqrt{4 \pi \rho} \ell_{T}}{\tau_{cool}}.
\end{equation}
The final strength of the generated small-scale field turns out to
be the same for both unstable zones. For the largest turbulent scale, 
$\ell_{T} =L \sim 1-3$ km, estimate (13) yields $B_{eq} \sim 3 \times 
10^{13}$ G if $\tau_{cool}$ is of the order of a few seconds. Using 
this estimate of $B_{eq}$, we can conclude that the strength of a 
large-scale poloidal field generated by the mean-field dynamo (equation
(8)) is in a good agreement with the observed magnetic fields of the 
majority of pulsars. For example, the generated poloidal field is
$\sim (1-2) \times 10^{13}$ G if the star rotates with the period
$ \sim 100$ ms. Note that extremely rapidly rotating PNSs ($P \sim
1$ ms) may possess a very strong poloidal field $\sim (3-6) \times
10^{14}$ G comparable to that of magnetars.

\acknowledgements 
VU thanks INFN (Catania) 
and Dipartimento di Fisica ad Astronomia, University of Catania, 
for financial support. AB thanks L.Rezzolla for important comments
on the manuscript.

{}

\end{document}